\documentclass[aip,
amsmath,amssymb,
 reprint]{revtex4-1}

\usepackage{graphicx}
\usepackage{dcolumn}
\usepackage{bm}

\usepackage[utf8]{inputenc}
\usepackage[T1]{fontenc}
\usepackage{mathptmx}
\usepackage{etoolbox}
\usepackage{amsmath} 
\usepackage{xcolor}

\usepackage{xr}
\makeatletter
\def\@email#1#2{%
 \endgroup
 \patchcmd{\titleblock@produce}
  {\frontmatter@RRAPformat}
  {\frontmatter@RRAPformat{\produce@RRAP{*#1\href{mailto:#2}{#2}}}\frontmatter@RRAPformat}
  {}{}
}%
\makeatother
\begin{document}

\preprint{AIP/123-QED}

\title[]{Thermal Conductivity and Temperature-Induced Band Gap Renormalization in Crystalline and Amorphous Ga$_2$O$_3$}

\author{Rustam Arabov*}
 \affiliation{Applied AI Institute, Moscow, Russia}
\email{rust.arabov@gmail.com}
 
\author{Jiaxuan Li}%
\affiliation{Institute of Nuclear and New Energy Technology, Tsinghua University, 100084, Beijing, China}%

\author{Xiaotong Chen}
\affiliation{Institute of Nuclear and New Energy Technology, Tsinghua University, 100084, Beijing, China}

\author{Nikita Rybin}
\affiliation{Applied AI Institute, Moscow, Russia}%
\affiliation{Digital Materials LLC, Bolshoi bulvar 30, build.1, 121205, Moscow, Russia}%
\affiliation{Moscow Engineering Physics Institute, Kashirskoe Highway 31, Moscow, 115409, Russia}

\author{Alexander Shapeev}
\affiliation{Applied AI Institute, Moscow, Russia}%
\affiliation{Digital Materials LLC, Bolshoi bulvar 30, build.1, 121205, Moscow, Russia}%

\date{\today}

\begin{abstract}

The lattice thermal conductivity (LTC) and electron–phonon interactions in crystalline and amorphous gallium oxide are herein determined by coupling a machine-learned interatomic potential, namely the moment tensor potential (MTP) model, to first-principles calculations. Crystalline $\beta$-Ga$_2$O$_3$ exhibits a substantial band gap renormalization (BGR) of $\sim$0.45~eV at 700~K, with $\sim$0.2~eV caused by zero-point BGR. The computed temperature dependence of BGR induced by classical nuclear motion in $\beta$-Ga$_2$O$_3$ is stronger than that in amorphous Ga$_2$O$_3$, with the difference in BGR reaching $\sim$0.18 eV at 900 K. Thermal transport calculations reveal that the LTC of amorphous Ga$_2$O$_3$ remains near $0.9$~W$\cdot$~m$^{-1}$$\cdot$K$^{-1}$ for temperatures between 300~K and 700~K, which is approximately an order of magnitude lower than that of crystalline $\beta$-Ga$_2$O$_3$. Overall, the presented framework provides a computationally tractable and reliable route for predicting properties of semiconductors (both crystalline and amorphous) under operating conditions relevant to microelectronics and optoelectronics.

\end{abstract}

\maketitle

Gallium oxide is among the most promising materials for applications in novel electronic and optoelectronic devices, which are rapidly developing nowadays. $\beta$-Ga$_2$O$_3$ is a wide band gap semiconductor applied in high-power electronic devices\cite{ga2o3_el_devices}, light detectors\cite{ga2o3_light_detectors}, and LEDs\cite{ga2o3_leds}. Amorphous gallium oxide can be used as a material for heat insulation coatings\cite{csanyi_ltc_agao,heat_coat_ga2o3}, gas sensors, photodetectors, and memory devices\cite{agao_gas_sens_mem_phot}. The application potential of Ga$_2$O$_3$ is determined by its thermal transport and electronic properties, which are influenced by lattice vibrations, or phonons. Phonons transfer heat, defining the lattice thermal conductivity (LTC), which is the main part of thermal conductivity in semiconductors\cite{ltc_overview}. Lattice vibrations also affect the electronic band structure through electron-phonon coupling, leading to band gap renormalization (BGR)\cite{zacharias2020fully}. These effects of lattice vibrations on thermal transport and electronic properties can be investigated by several \textit{ab initio} DFT-based computational methods. For example, the approach based on the solution of the Boltzmann transport equation (BTE) for phonons\cite{ltc_bte, ltc_bte_phono3py, ltc_bte_tdep} and the Green-Kubo method\cite{green-kubo, knoop} are applicable for calculating LTC of solids. Phonon-induced BGR can be calculated using the Allen-Heine-Cardona theory\cite{bgr_ahc}.

However, these approaches are computationally prohibitive for complex materials when using \textit{ab initio} methods to calculate interatomic energies and forces. Examples of such complex materials are crystalline ($\beta$-Ga$_2$O$_3$) and amorphous Ga$_2$O$_3$ (a-Ga$_2$O$_3$). The complexity of $\beta$-Ga$_2$O$_3$ arises from its low-symmetry crystal lattice. The primitive cell of this material consists of 10 atoms, and its lattice belongs to space group C2/m. Thus, LTC calculation within the BTE-based approach for $\beta$-Ga$_2$O$_3$ requires computing interatomic forces for 11625 120-atom supercells. Amorphous a-Ga$_2$O$_3$ is also a complex material because it is completely asymmetric. To compute its LTC, methods based on long-term molecular dynamics simulations must be used. Direct DFT calculations are extremely expensive for obtaining the LTC of both $\beta$-Ga$_2$O$_3$ and a-Ga$_2$O$_3$. Nevertheless, machine-learned interatomic potentials, such as moment tensor potentials (MTPs)\cite{mlip}, provide a way to overcome this limitation. MTPs can calculate interatomic energies and forces with accuracy close to DFT, but much more efficiently. Thus, using MTPs instead of DFT within the aforementioned approaches for obtaining LTC and BGR can make the calculations far less expensive\cite{rybin_mtp_ga2o3, yang2026neuroevolution}.

In this work, we investigate heat transport and electron-phonon coupling in $\beta$-Ga$_2$O$_3$ and a-Ga$_2$O$_3$. Using MTPs to compute interatomic energies and forces, we study BGR induced by lattice vibrations in these materials and calculate LTC for a-Ga$_2$O$_3$ with the Green-Kubo approach. The fundamental gaps computed from the electronic densities of states are considered as band gaps in our work. Our results (BGR in $\beta$-Ga$_2$O$_3$ and LTC of a-Ga$_2$O$_3$) show good agreement with experimental data\cite{onuma_bgr_T_exp, sturm_bgr_T_exp, mock_bgr_T_exp, lee_beta_ga2o3_bgr}.

The MTPs used in this work were trained on DFT data. DFT computations were performed using the PAW method\cite{PAW} and the PBE\cite{pbe} parametrization of the generalized gradient approximation\cite{gga} in VASP\cite{vasp} (the effect of the exchange-correlation functional on BGR calculation results is discussed in the supplementary material). The 3d electrons of Ga were explicitly included in the corresponding pseudopotential. The plane-wave basis cutoff energy and the electronic self-consistency threshold for all DFT calculations were set to 500 eV and $10^{-6}$ eV, respectively.

The procedure for MTP training and validation for $\beta$-Ga$_2$O$_3$ was as follows. An MTP of level 18 (325 parameters) was initially trained on the set of configurations from the work of Rybin and Shapeev\cite{rybin_mtp_ga2o3}. Then, the active learning algorithm implemented in the MLIP-2 package\cite{podryabinkin2017active, mlip} was used to additionally train this MTP. The accuracy of the MTP was validated by comparing the results of unit cell relaxation, interatomic forces and phonon band structure calculations with this MTP and with DFT. The phonon band structures were obtained using Phonopy\cite{phonopy}, based on the interatomic force constants computed with the MTP and DFT. Good agreement between the MTP and DFT results was achieved.

\begin{figure*}[t!]
\includegraphics[width=0.9\linewidth]{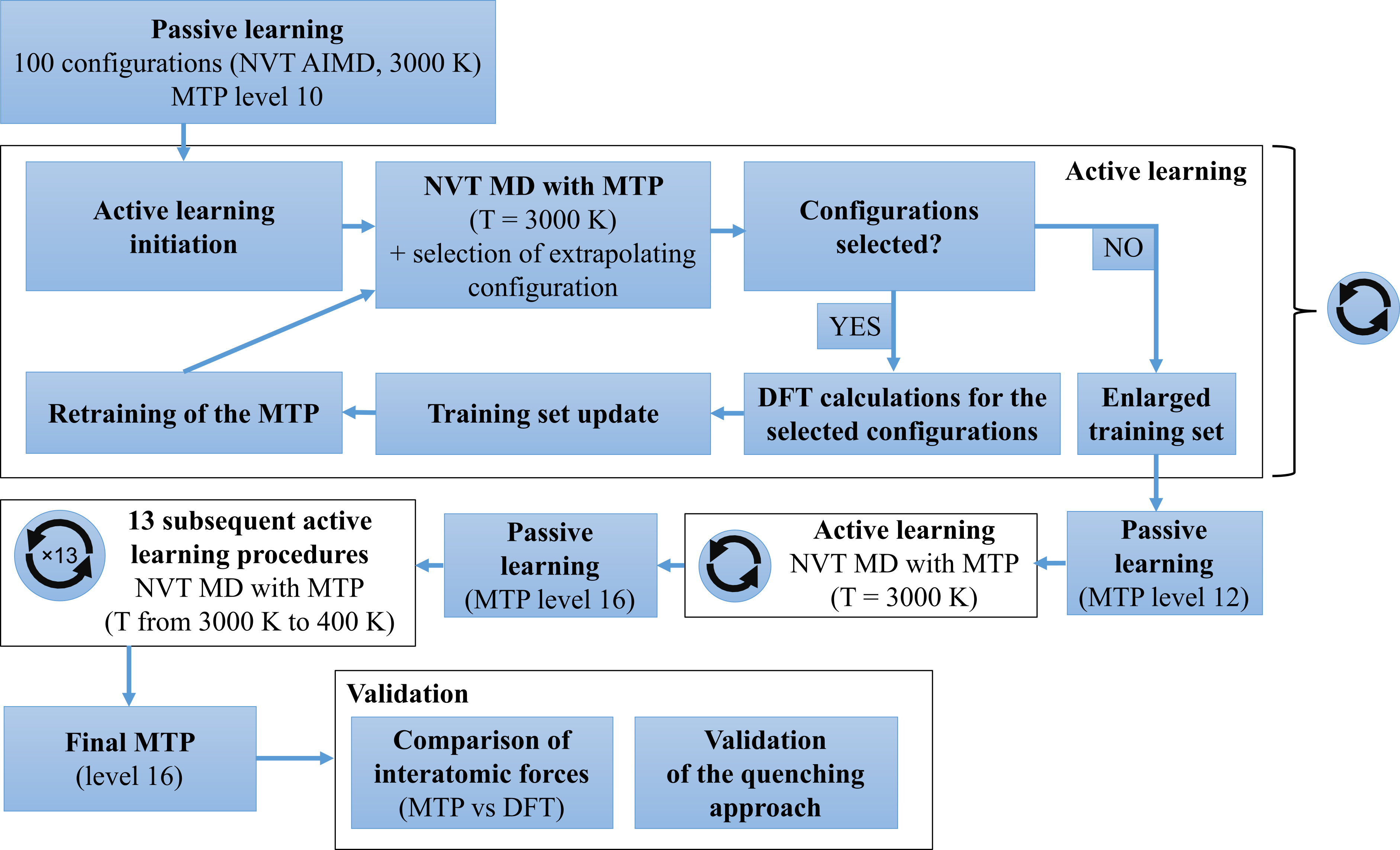}
\caption{Workflow of MTP training and validation for a-$\text{Ga}_2\text{O}_3$. The active learning procedure, indicated by the symbol of two circling arrows, is detailed in the block ``Active learning''. This procedure is the same for all blocks containing this symbol; the only difference lies in the conditions of the MD simulation.}
\label{fig:workflow_agao}
\end{figure*}

The MTP for a-Ga$_2$O$_3$ was trained and validated as shown in Fig. \ref{fig:workflow_agao}. First, the initial cell of amorphous Ga$_2$O$_3$ was generated by means of random packing. Second, an \textit{ab initio} molecular dynamics (AIMD) simulation at 3000 K was run starting from this initial cell. The snapshots from this AIMD trajectory were used to fit the initial MTP of level 10 (114 parameters). Then, the MTP was fitted using active learning implemented in the MLIP-2 package\cite{podryabinkin2017active, mlip}. The training set generated after this active learning procedure was applied to train the MTP of level 12 (127 parameters). Subsequent active learning of this MTP yielded an expanded training set, which was used to fit the MTP of level 16 (222 parameters). This MTP of level 16 was trained with active learning over a temperature range of 3000–400 K to model the quenching process. The resulting MTP of level 16 was validated by comparing the results of interatomic forces prediction for a-Ga$_2$O$_3$ performed with the MTP and DFT. The MTP-based quenching approach applied to create physically realistic cells of a-Ga$_2$O$_3$ was validated by comparing the pair distribution functions for 15 quenched cells with literature data\cite{sun2020matching, csanyi_ltc_agao}. Details on the training and validation of the MTP for $\beta\text{-Ga}_2\text{O}_3$ and a-Ga$_2$O$_3$ are provided in the supplementary material. Overall, we concluded that the MTPs described above allowed us to correctly model the interatomic interactions in $\beta$-Ga$_2$O$_3$ and a-Ga$_2$O$_3$. 

After validating the accuracy of the MTPs, we investigated the properties of gallium oxide using these potentials. We started by computing BGR in $\beta$-Ga$_2$O$_3$ at temperatures of 0 K, 100 K, 300 K, 500 K, 700 K, and 900 K. For each temperature, sets of 120-atom $\beta$-Ga$_2$O$_3$ supercells with displaced atoms were created using several approaches. To describe nuclear motion in the harmonic approximation, the method based on the superposition of normal phonon modes implemented in the HIPHIVE package\cite{hiphive} was applied, utilizing the second-order force constants computed with the MTP. This method is also referred to as harmonic sampling. At each temperature, 25 supercells with displaced atoms were created using both classical and quantum statistics, the band gaps $E_{\rm g}(T)$ were computed for these supercells with DFT and averaged. We also calculated $E_{\rm g}(0)$ for the undisturbed 120-atom supercell. The ${\rm BGR}(T)$ values were calculated by subtracting $E_{\rm g}(0)$ from the averaged temperature-dependent band gaps. To estimate the contribution of quantum nuclear effects to electron-phonon coupling in $\beta$-Ga$_2$O$_3$, we compared the temperature-dependent BGR calculated using classical (${\rm BGR}^{\rm HS}_{\rm class}(T)$) and quantum (${\rm BGR}^{\rm HS}_{\rm quan}(T)$) statistics for characterizing nuclear motion. In the classical case, the temperature dependence of BGR was fitted with a linear function (${\rm BGR}^{\rm HS}_{\rm class}(T) = a\cdot T$). The ${\rm BGR}^{\rm HS}_{\rm quan}(T)$ dependence was fitted with the modified Varshni's expression:

\begin{equation}\label{eq:varshni}
    {\rm BGR}^{\rm HS}_{\rm quan}(T) = {\rm ZPR} - \frac{\alpha T^2}{T+\beta}.
\end{equation}
Here $\rm ZPR$, $\alpha$, and $\beta$ are fitting parameters ($\rm ZPR$ is zero-point band gap renormalization), and $T$ is temperature. The mathematical details on the BGR(T) calculation and Varshni's expression\cite{varshni} modification for fitting the ${\rm BGR}^{\rm HS}_{\rm quan}(T)$ are given in the supplementary material.

\begin{figure}[h!]
\includegraphics[width=0.75\linewidth]{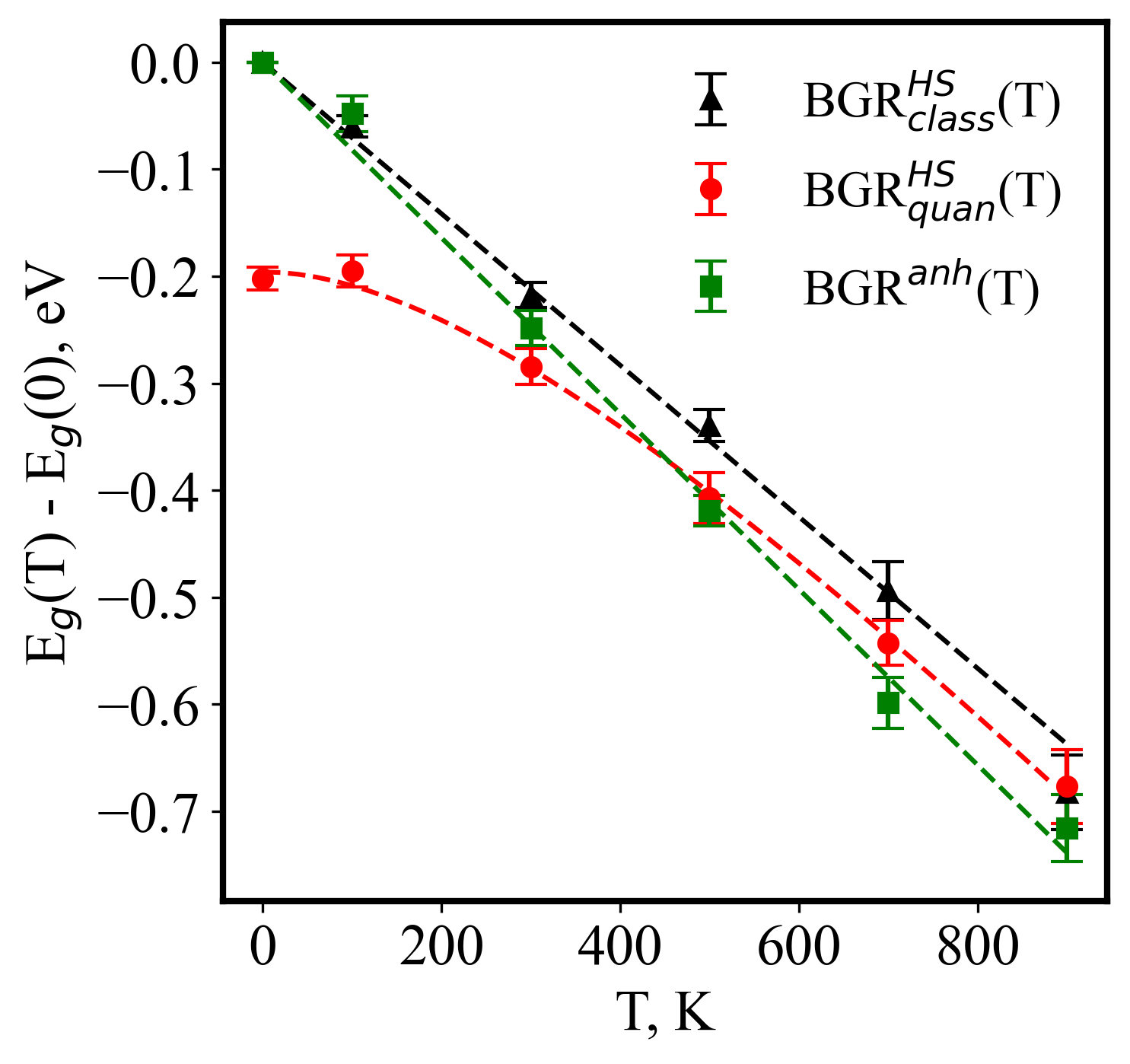}
\caption{Band gap renormalization in $\beta$-Ga$_2$O$_3$ computed using  configurations generated by means of harmonic sampling (using classical statistics -- ${\rm BGR}^{\rm HS}_{\rm class}(T)$, using quantum statistics -- ${\rm BGR}^{\rm HS}_{\rm quan}(T)$) and taking anharmonic effects into account -- ${\rm BGR}^{\rm anh}(T)$.}
\label{fig:bgr_clas_quan}
\end{figure}

The results of the ${\rm BGR}^{\rm HS}_{\rm class}(T)$ and ${\rm BGR}^{\rm HS}_{\rm quan}(T)$ calculations are shown in Fig.\ \ref{fig:bgr_clas_quan}. The error bars in all ${\rm BGR}(T)$ graphs below represent the standard errors of the mean for the BGR values.

According to the data shown in Fig.\ \ref{fig:bgr_clas_quan}, the ZPR in $\beta$-Ga$_2$O$_3$ is 0.20$\pm$0.01 eV. This value is considerable and close to the ZPR in materials such as SiC (ZPR 0.18 eV), GaN (ZPR 0.18 eV), and ZnO (ZPR 0.16 eV)\cite{gonze_bgr_database}.

Next, we analyzed the influence of lattice dynamics anharmonicity on BGR in this material. First, we computed BGR taking into account the anharmonicity of nuclear motion using samples from NVT MD trajectories. Classical NVT MD simulations lasting 100 ps with a 1 fs timestep were performed with the supercell consisting of 120 atoms at each temperature of interest. 35 samples separated by 1 ps intervals were taken from each MD trajectory for further BGR calculations. MD simulations with the MTP were run using the LAMMPS software\cite{lammps} with the LAMMPS-MLIP interface\cite{mlip-lammps-GK}.

Second, the effect of lattice thermal expansion on BGR was added to ${\rm BGR}^{\rm MD}(T)$ as follows. Unit cells with volumes corresponding to the temperatures of 100 K, 300 K, 500 K, 700 K, and 900 K were created. The dependence of the unit cell volume on temperature was taken from experimental data reported in the literature\cite{orlandi_LTE_exp, villora_LTE_exp}. Single-point DFT calculations were performed for these unit cells. The differences between the band gaps of the unit cells with temperature-dependent volumes ($E^{\rm VT}_g(T)$) and the relaxed unit cell of $\beta$-Ga$_2$O$_3$ ($E^{\rm VT}_g(0)$) were calculated and added to the ${\rm BGR}^{\rm MD}(T)$ values:

\begin{equation}\label{eq:md+lte}
{\rm BGR}^{\rm anh} = {\rm BGR}^{\rm MD}(T) + E^{\rm VT}_{\rm g}(T)- E^{\rm VT}_{\rm g}(0).
\end{equation}

Here ${\rm BGR}^{\rm anh}(T)$ is the dependence of BGR on temperature calculated taking into account two anharmonic effects (anharmonicity of nuclear motion and lattice thermal expansion)  in $\beta$-Ga$_2$O$_3$ over the temperature range from 0 K to 900 K.

As shown in Fig.\ \ref{fig:bgr_clas_quan}, consideration of nuclear motion anharmonicity and lattice thermal expansion led to a small change in the temperature dependence of BGR. Therefore, the contribution of anharmonic effects to electron-phonon coupling is not significant in $\beta$-Ga$_2$O$_3$.

\begin{figure}[h!]
\includegraphics[width=0.75\linewidth]{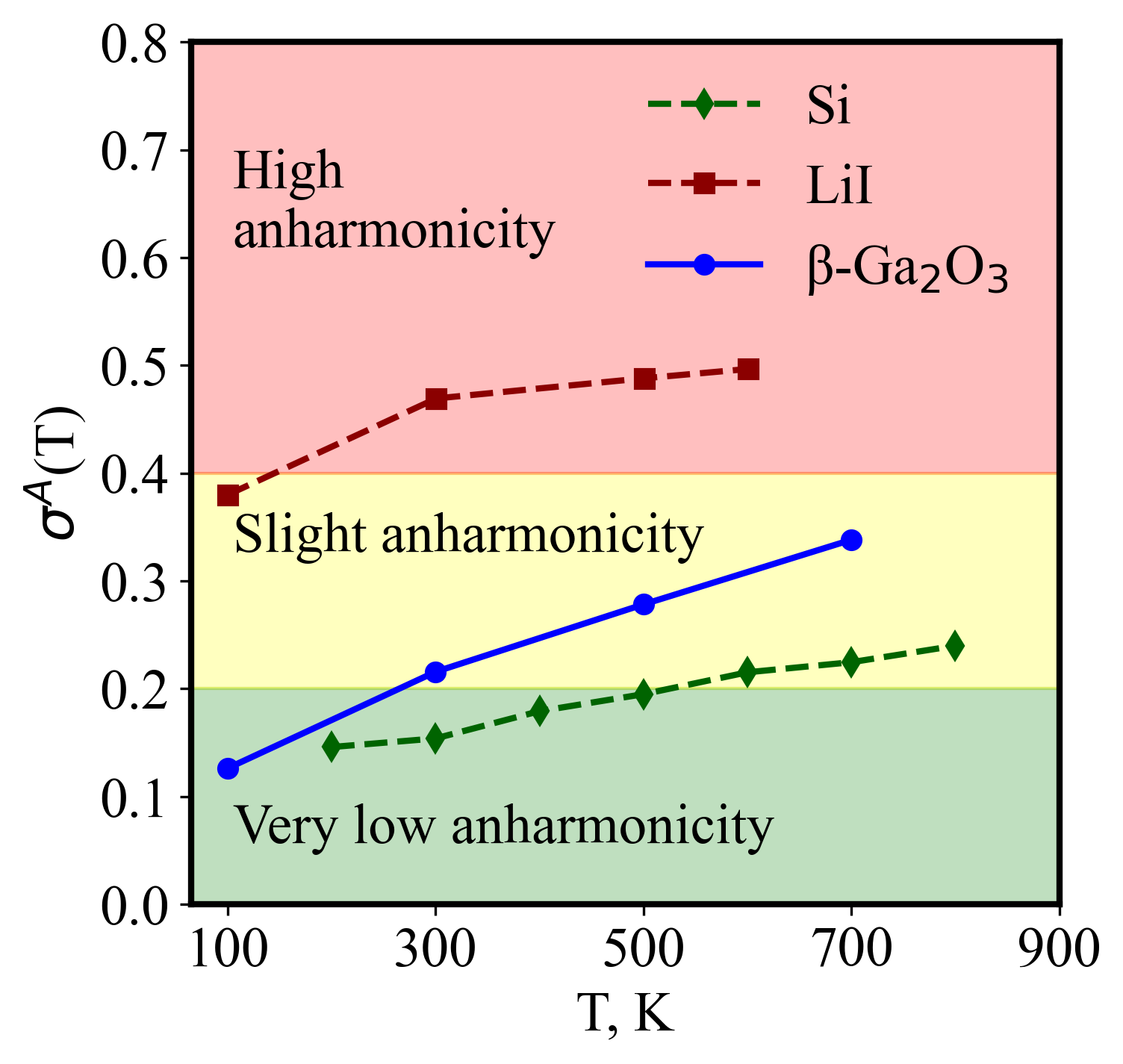}
\caption{Anharmonicity measure $\sigma^\text{A}$ as a function of temperature for $\beta$-Ga$_2$O$_3$ (present work, obtained from MD simulations), Si\cite{knoop2020anharmonicity}, and LiI\cite{rybin_disser}.}
\label{fig:anh_meas}
\end{figure}

The insignificant influence of lattice dynamics anharmonicity on BGR in $\beta$-Ga$_2$O$_3$ agrees with the values of the anharmonicity measure\cite{knoop2020anharmonicity} $\sigma^\text{A}$(T) for this material. $\sigma^\text{A}$(T) quantifies the ratio of forces arising from anharmonic interactions to the total interatomic forces and can be used to distinguish harmonic materials from anharmonic ones. In Fig.\ \ref{fig:anh_meas}, the temperature dependencies of $\sigma^\text{A}$ for Si, LiI, and $\beta$-Ga$_2$O$_3$ are compared. Si and LiI are examples of harmonic and anharmonic materials, respectively. According to the literature data\cite{knoop2020anharmonicity, rybin_disser}, typical anharmonicity measure values for very harmonic materials are $\sigma^\text{A} < 0.2$, and for strongly anharmonic materials $\sigma^\text{A}$ is close to 1. For $\beta$-Ga$_2$O$_3$, $\sigma^\text{A}$ was higher than 0.2 at $T > 300$ K and did not become close to 1 in the temperature range from 100 K to 700 K (the highest value of $\sigma^\text{A}$ in this temperature range was 0.34). Therefore, this material can be considered slightly anharmonic, which agrees with the small effect of anharmonicity on BGR. Notably, the increase in anharmonic contributions to interatomic forces with temperature was consistent with the larger difference between ${\rm BGR}^{\rm anh}(T)$ and ${\rm BGR}^{\rm HS}_{\rm class}(T)$ at higher temperatures (see Fig.\ \ref{fig:bgr_clas_quan}).

After considering the lattice dynamics anharmonicity, ${\rm BGR}^{\rm anh+Q}(T)$ was computed taking into account quantum nuclear and anharmonic effects together:
\begin{equation}
\begin{split}
    {\rm BGR}^{\rm anh+Q}(T) = {\rm BGR}^{\rm anh}(T)\\ + {\rm BGR}^{\rm HS}_{\rm quan}(T) - {\rm BGR}^{\rm HS}_{\rm class}(T)
\end{split}
\end{equation}

\begin{figure}[h!]
\includegraphics[width=0.75\linewidth]{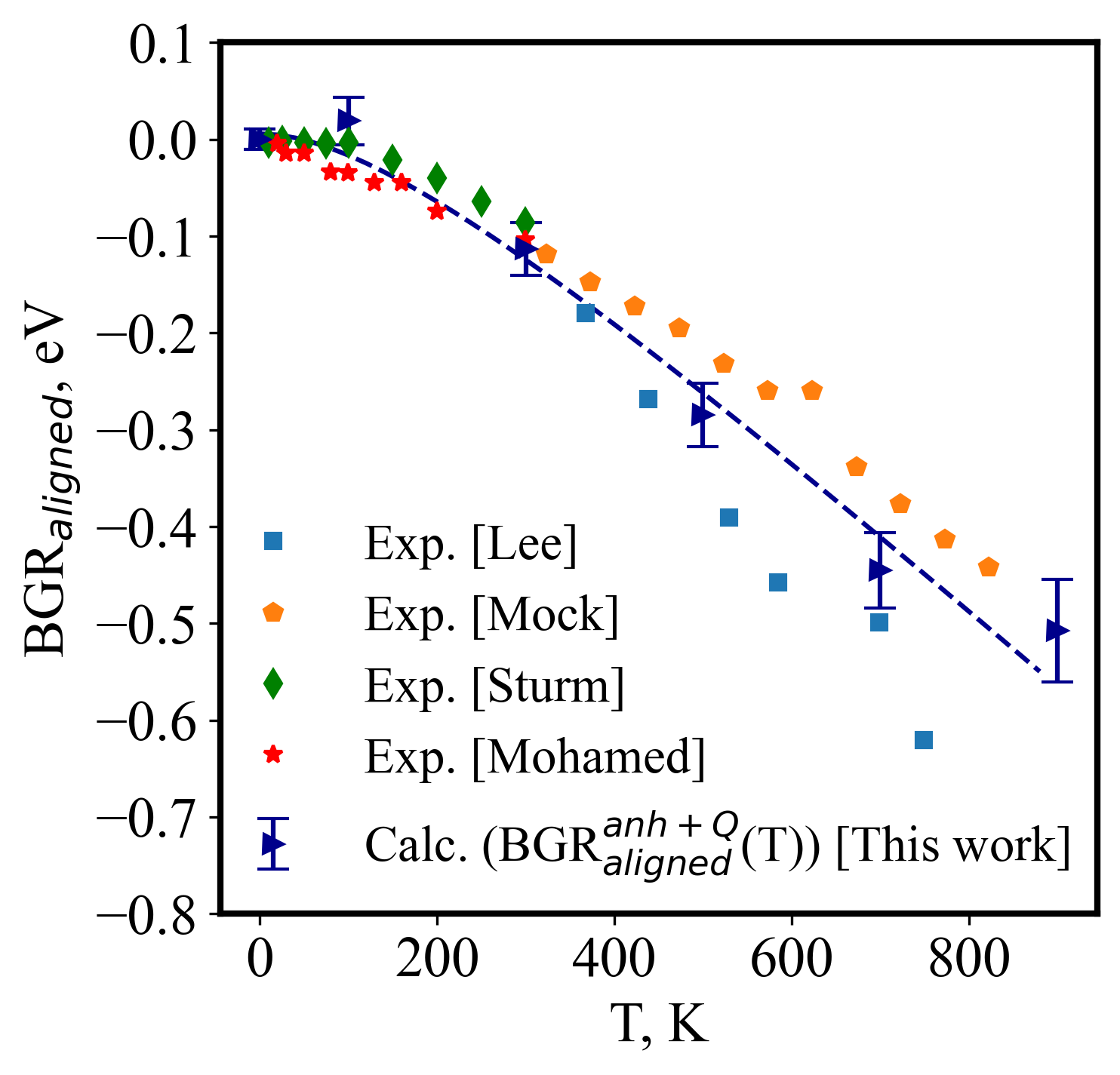}
\caption{Band gap renormalization ${\rm BGR}^{\rm anh+Q}_{\rm aligned}(T)$ compared with experimental data\cite{lee_beta_ga2o3_bgr,mock_bgr_T_exp,sturm_bgr_T_exp,mohamed_exp_bgr}.}
\label{fig:bgrq-vs-exp}
\end{figure}

These ${\rm BGR}^{\rm anh+Q}(T)$ calculations showed a significant change in the band gap with temperature in crystalline $\beta$-Ga$_2$O$_3$ ($E^{\rm anh + Q}_{\rm g}(\text{700 K}) - E^{\rm anh + Q}_{\rm g}(\text{0 K}) \approx 0.45$ eV). This result was in good agreement with experimental data\cite{onuma_bgr_T_exp, sturm_bgr_T_exp, mock_bgr_T_exp, mohamed_exp_bgr}, as shown in Fig.\ \ref{fig:bgrq-vs-exp}. To compare our predictions with the experimentally measured BGR values, the experimental and theoretical BGR(T) dependencies were aligned. The alignment procedure is described in the supplementary material.

Additionally, the effect of long-range electrostatics on BGR in $\beta$-Ga$_2$O$_3$ was studied. The mathematical details and results of these calculations are given in the section ``Effect of Fr$\Ddot{\text{o}}$hlich polar coupling on BGR in $\beta\text{-Ga}_2\text{O}_3$'' in the supplementary material.

After studying electron-phonon coupling in crystalline $\beta$-Ga$_2$O$_3$, we moved to the investigation of BGR in amorphous gallium oxide. To compute the zero-temperature band gap of a-Ga$_2$O$_3$, 15 configurations consisting of 200 atoms were generated by means of random packing and quenched. The band gaps obtained for these 15 configurations were averaged, and the average value was considered as the band gap $E^a_g(0)$ of a-Ga$_2$O$_3$ at 0 K (without taking into account zero-point vibrations of nuclei). The details of how the number of samples for band gap averaging was chosen are given in the supplementary material. After computing $E^a_g(0)$, the band gap values were calculated at non-zero temperatures. For this, 3 samples from the set of configurations used to obtain $E^a_g(0)$ were kept at 300 K, 400 K, 500 K, and 900 K for 15 ps (NVT MD with the MTP, 1 fs timestep). Single-point DFT calculations were performed for 15 samples (separated by 1 ps intervals) from each MD trajectory. For each temperature, the obtained band gap values were averaged over all samples extracted from the 3 MD trajectories related to that temperature. The BGR values for a-Ga$_2$O$_3$ were calculated by subtracting $E^a_g(0)$ from the average band gaps computed at non-zero temperatures. 

\begin{figure}[h!]
\includegraphics[width=0.75\linewidth]{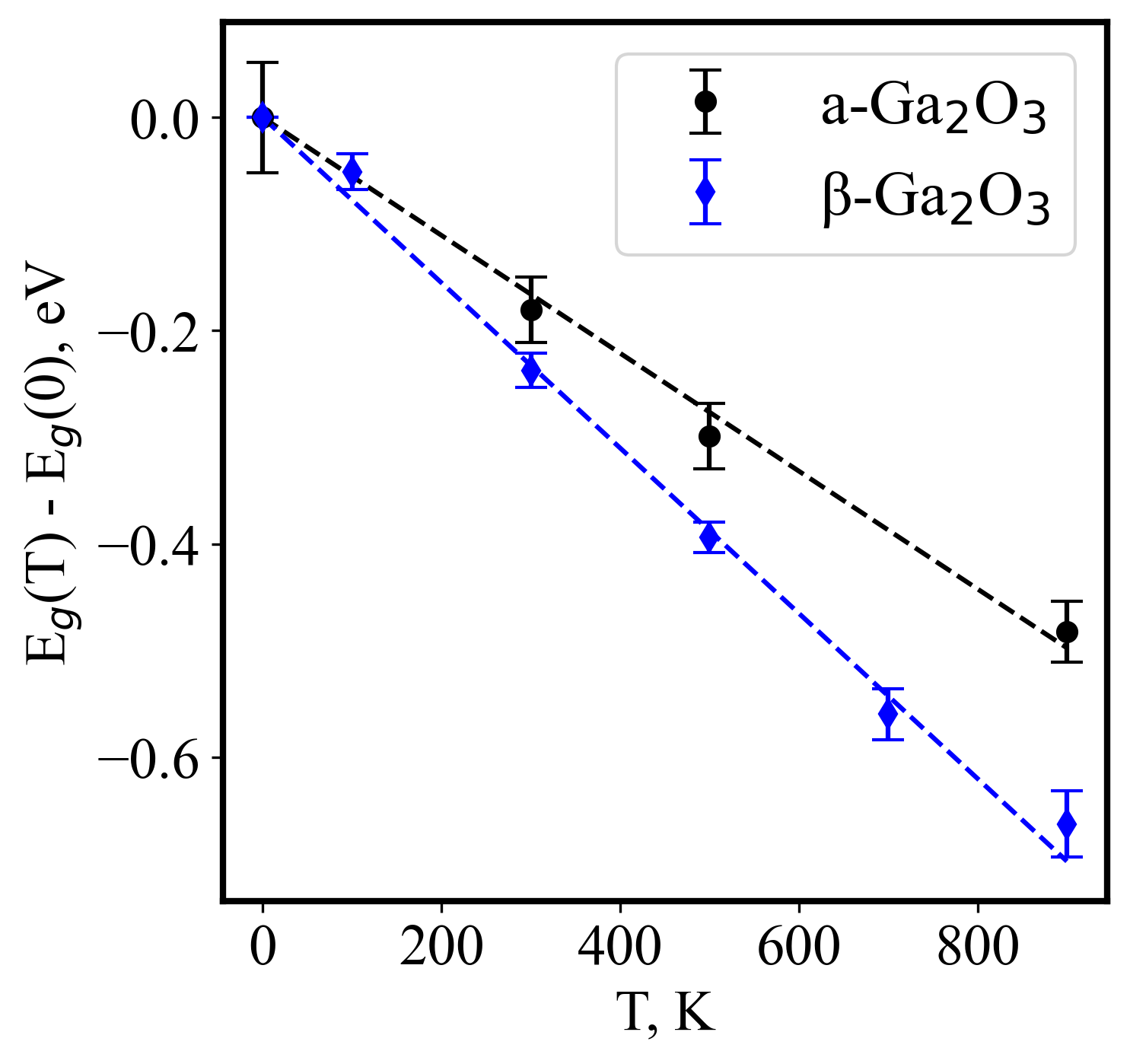}
\caption{Comparison of ${\rm BGR}^{\rm MD}(T)$ values for a-Ga$_2$O$_3$ and $\beta$-Ga$_2$O$_3$ calculated using configurations taken from NVT MD trajectories.}
\label{fig:bgr-agao-vs-beta}
\end{figure}

As shown in Fig.\ \ref{fig:bgr-agao-vs-beta}, at a fixed temperature the absolute BGR values for $\beta$-Ga$_2$O$_3$ were higher than those for amorphous gallium oxide. From this, we concluded that amorphization of Ga$_2$O$_3$ led to a weakening of the dependence of BGR on temperature.

In addition, according to the data given in Fig.\ \ref{fig:bgr-agao-vs-beta}, both crystalline and a-Ga$_2$O$_3$ exhibit significant band gap renormalization with temperature (${\rm BGR}^{\rm MD}(900~\text{K}) = 0.66 \pm 0.03$ eV and ${\rm BGR}^{\rm MD}(900~\text{K})=0.48\pm 0.03$ eV for $\beta$-Ga$_2$O$_3$ and a-Ga$_2$O$_3$, respectively). Therefore, it is important to consider the effect of electron-phonon coupling on the band gap when calculating the electronic properties of Ga$_2$O$_3$.

After analyzing BGR in a-Ga$_2$O$_3$, we moved on to the characterization of heat transport in this material. The method based on the Green-Kubo formula\cite{lammps_GK, green-kubo, knoop} was applied to calculate the LTC of a-Ga$_2$O$_3$. LTC values for a-Ga$_2$O$_3$ were obtained at temperatures of 300 K, 400 K, 500 K, and 700 K. At each temperature, 20 independent MD simulations were carried out with the MTP, and the LTC values computed from the 20 MD trajectories were averaged to obtain the final temperature dependence of LTC.

\begin{figure}[h!]
\includegraphics[width=\linewidth]{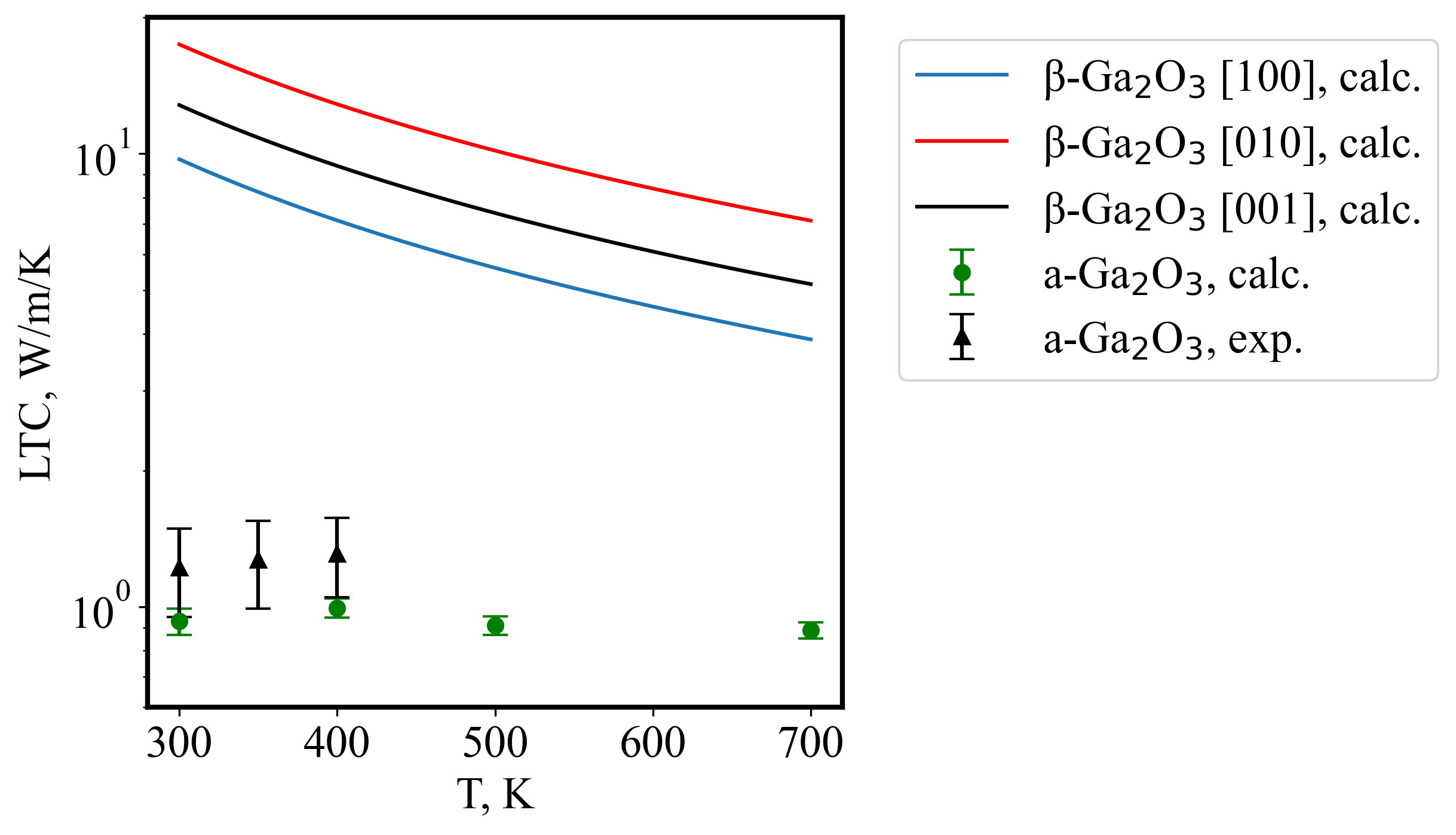}
\caption{LTC of a-Ga$_2$O$_3$ and $\beta$-Ga$_2$O$_3$: theoretical LTC values for a-Ga$_2$O$_3$ (a-Ga$_2$O$_3$, calc) -- calculated in the present work,  experimental thermal conductivity of a-Ga$_2$O$_3$ (a-Ga$_2$O$_3$, Exp.) -- from the work of Liu et al.\cite{csanyi_ltc_agao}, and theoretical LTC values for $\beta$-Ga$_2$O$_3$ -- from the work of Rybin and Shapeev\cite{rybin_mtp_ga2o3}.}
\label{fig:ltc-agao-vs-exp}
\end{figure}

The results of the LTC calculations for a-Ga$_2$O$_3$ were compared with experimental data\cite{csanyi_ltc_agao}. As shown in Fig.\ \ref{fig:ltc-agao-vs-exp}, the calculated values were slightly lower than the experimental thermal conductivity. However, our calculations successfully captured the weak dependence of LTC on temperature observed experimentally across the studied temperature range. The discrepancy between the theoretical and experimental LTC values can be attributed to finite-size effects of the simulation system and the composition of the sample used in the experiment\cite{csanyi_ltc_agao}. A detailed explanation of the reasons for this discrepancy is given in the supplementary material.

In addition, the LTC of a-Ga$_2$O$_3$ was compared with the data on $\beta$-Ga$_2$O$_3$ from the work of Rybin and Shapeev\cite{rybin_mtp_ga2o3}, where an MTP was also used for computing interatomic forces during the LTC calculations. The theoretical LTC of a-Ga$_2$O$_3$ was 6--14 times lower than that of $\beta$-Ga$_2$O$_3$ in the temperature range from 300 K to 700 K, which is also consistent with experimental data\cite{csanyi_ltc_agao}. The reason for the significant LTC difference between crystalline and amorphous Ga$_2$O$_3$ is as follows. In $\beta$-Ga$_2$O$_3$, thermal transport is dominated by propagating phonon modes. In contrast, the absence of long-range structural order in amorphous Ga$_2$O$_3$ strongly localizes lattice vibrations. This localization traps thermal energy, which results in lower LTC values for a-Ga$_2$O$_3$ compared to $\beta$-Ga$_2$O$_3$. More details on the LTC calculations for a-Ga$_2$O$_3$ are given in the supplementary material.

Overall, electron-phonon coupling and thermal transport properties were systematically investigated for crystalline and amorphous gallium oxide. The calculations were done using DFT and machine-learned moment tensor interatomic potentials. Based on the analysis of the obtained results, the following conclusions were made.

First, we found that zero-point nuclear vibrations made a notable contribution to BGR in $\beta$-Ga$_2$O$_3$ ($\rm ZPR = 0.20 \pm 0.01$ eV), whereas anharmonic effects had only a minor influence on BGR at temperatures from 0 K to 900 K. Second, we found that amorphization weakened the temperature dependence of BGR. However, significant BGR was observed in both crystalline and amorphous phases: ${\rm BGR}^{\rm MD}(900~\text{K})$ is $0.66 \pm 0.03$ eV for $\beta$-Ga$_2$O$_3$ and $0.48 \pm 0.03$ eV for a-Ga$_2$O$_3$. Third, we showed that the LTC of amorphous Ga$_2$O$_3$ was approximately an order of magnitude lower than that of its crystalline $\beta$-phase counterpart. This was attributed to the loss of long-range atomic order in the amorphous state.

See the supplementary material for a detailed description of MTP training, amorphous Ga$_2$O$_3$ structures generation and quenching, BGR analysis, and LTC computations with the Green-Kubo method.

R.A., N.R., and A.S. acknowledge funding from the Russian Science Foundation (grant number 23-13-00332-P). J.L. and X.C. acknowledge financial support from the National S$\text{\&}$T Major Project (Grant No. ZX060901), Research Fund for Young Talents CNNC, and they also thank the Supercomputing Network and the High Performance Computing Center of Tsinghua University for providing computational resources.

\bibliography{ref}

\end{document}